\RequirePackage{fix-cm}
\documentclass[twocolumn]{svjour3}          
\smartqed  
\usepackage{graphicx}
\journalname{Journal of Radioanalytical Nuclear Chemistry}
\begin{document}

\title{Cross-section measurement of some deuteron induced reactions on $^{160}$Gd for possible production of the therapeutic radionuclide $^{161}$Tb
}


\author{F. T\'ark\'anyi \and  A. Hermanne \and S. Tak\'acs \and  F. Ditr\'oi \and J. Csikai\and A.V. Ignatyuk
}


\institute{F. Ditr\'oi F. T\'ark\'anyi \and S. Tak\'acs \and J. Csikai \at
              Institute of Nuclear Research of the Hungarian Academy of Sciences \\
              Tel.: +36-52-509251\\
              Fax: +36-52-416181\\
              \email{ditroi@atomki.hu}           
           \and
           A. Hermanne \at
              Cyclotron Laboratory, Vrije Universiteit Brussel (VUB), Brussels, Belgium
           \and
              A.V. Ignatyuk \at
            Institute of Physics and Power Engineering (IPPE), Obninsk 249020, Russia  
}

\date{Received: 2013 / Accepted: 2013}

\maketitle

\begin{abstract}
The radionuclide $^{161}$Tb (T$_{1/2}$=6.89 d) is potentially important for internal radiotherapy. It is generally produced through the $^{160}$Gd(n,$\gamma$)$^{161}$Gd $\rightarrow$ $^{161}$Tb route at research reactors. In this work the possibility of its production at a cyclotron was investigated. Determination of the excitation function of the $^{160}$Gd(d,x)$^{161}$Tb production route and that of the disturbing $^{160}$Gd(d,2n)\-$^{160}$Tb side reaction was done over the deuteron energy range up to 50 MeV using the stacked-foil technique and high-resolution  $\gamma$-ray spectrometry. A comparison of this production route with the established (n,$\gamma$) reaction at a nuclear reactor is made.
\end{abstract}

\section{Introduction}
\label{intro}
Radioisotopes are essential for a variety of applications in medicine as diagnosis by scintigraphy and treatment of various diseases by internal radiotherapy. There is a rapid growth in the use of radionuclides for treatment of cancer in nuclear medicine paralleled by an increase of the diversity of the used radioisotopes. 
$^{161}$Tb has appropriate decay properties for use in cancer therapy \cite{1,2,3,4}.
\begin{itemize}
\item   Its half-life of 6.9 days - long enough to allow transport to hospitals but short enough to avoid long term issues of waste handling after excretion from the patient.
\item   Emits low-energy $\beta^-$- particles and low-energy Auger-electrons what results in a short cytotoxic range with minimal collateral damage to healthy tissue.
\item    Emits a small amount of  $\gamma$-radiation - enough to detect exactly where the radioisotope has been delivered to.
\end{itemize}
It has similar decay properties, biochemistry and in vivo property as the commercially available $^{177}$Lu radioisotope, but it emits more low-energy (Auger) electrons, meaning it could be better in treating smaller tumors.
The radionuclide $^{161}$Tb can be obtained in n.c.a. (non-carrier-added) form by irradiating highly enriched, massive $^{160}$Gd targets with intense neutron beams in nuclear research reactors relying on the indirect $^{160}$Gd(n,$\gamma$)\-$^{161}$Gd$\rightarrow$$^{161}$Tb production route [1]. This approach is similar to the indirect production route used for $^{177}$Lu. The thermal neutron cross-section of the $^{160}$Gd(n,$\gamma$) reaction (1.5 barn) is however lower than the one for the $^{176}$Yb(n,$\gamma$)\-$^{176}$Yb$\rightarrow$ $^{177g}$Lu reaction (3 barn), therefore longer irradiation or higher flux is required for production of the same batch activity. 
The production possibility of using the photonuclear reaction $^{162}$Lu($\gamma$,p) on highly enriched $^{162}$Lu was discussed in \cite{5}. It will allow to produce $^{161}$Tb with rather high specifc activity and low radionuclidic impurity, but the cross-sections for ($\gamma$,p) reactions are known to be low. There is also a possibility to produce $^{161}$Tb through the ($\alpha$,x) reactions from natural or enriched gadolinium \cite{6}, but the excitation function reaches only 13 mb at 60 MeV.
To increase the availability of $^{161}$Tb, we have investigated the possible production by deuteron-induced reactions on Gd. No earlier investigations for this production route were found in the literature. We should mention from the beginning that similar investigations were done for $^{177}$Lu by Hermanne et al. \cite{7} and by Manenti et al. \cite{8} showing the advantage of the reactor production.
We have used for our investigation gadolinium targets with natural isotopic composition ($^{152}$Gd-0.2\%, $^{154}$Gd-2.18\%, $^{155}$Gd-14.8\%, $^{156}$Gd-20.47\%, $^{157}$Gd-15.65\%, $^{158}$Gd-24.84\%, $^{160}$Gd-21.86\%) bombarded with a 50 MeV deuteron beam. As $^{161}$Tb can only be produced through reactions on the highest mass isotope $^{160}$Gd, among the numerous activation products assessed in this experiment we can single out cumulative production of $^{161}$Tb, via the direct (d,n) production and by the decay of $^{161}$Gd (3.66 min,   $\beta^{-}$: 100 \% ). Results for the other investigated radionuclides will be published separately.

\section{Experimental technique and data analysis}
\label{sec:2}
The experimental techniques and data analysis were similar as described by us in the recent publications (c.f. \cite{9,10,11}). Here we present shortly the most important factors, and details, specific for this experiment.
A first stack (series 1), containing  $^{nat}$Gd (83.9  min), Sc target foils (105.27  min), aluminum degrader and monitor foils (27  $\mu$m), was irradiated at the Cyclone 90 cyclotron of the Université Catholique in Louvain la Neuve (LLN) with a 50 MeV incident energy deuteron beam (42 min, 121 nA). A second stack (series 2.) containing   $^{nat}$Gd (83.9  m) target foils, aluminum degrader (98 $\mu$m) and Ti monitor foils (11 $\mu$m), was irradiated at the CGR 560 cyclotron of the Vrije Universiteit Brussel (VUB) with a 21 MeV incident energy deuteron beam (60 min, 110 nA).
The activity produced in the targets and monitor foils was measured non-destructively (without chemical separation) using a high resolution HPGe  $\gamma$-ray spectrometer. Three series of measurements at different detector-target distances were done starting about 2 h, 20 h and 200 h after EOB. The decay data were taken from the online database NuDat2 \cite{12}.
Effective beam intensities and the energy scale were determined by using the excitation functions of the $^{24}$Al(d,x)$^{22,24}$Na and $^{nat}$Ti(d,x)$^{48}$V reactions \cite{13} simultaneously re-measured over the whole energy range. The figure of the re-measured excitation function in comparison with the IAEA recommended data \cite{9,13} can be found in our earlier publication on Sc, irradiated in the same experiment \cite{11}.
Concerning the two related isotopes ($^{161}$Tb and its $^{161}$Gd parent) we could not assess the short-lived $^{161}$Gd (T$_{1/2}$ =3.66 min), which has decayed out at the moment of the first series of measurements. Only the $^{161}$Tb  $\gamma$-lines were identified. The 57.1 keV  $\gamma$-line of $^{161}$Tb (see Table 1) was used for the cross-section calculations, as the 74.6  $\gamma$-line overlaps with the lead KX-rays and the 49 keV  $\gamma$-line is contaminated with signal from the Tb, Gd and Eu KX-rays. At 25.6 keV  $\gamma$-ray energy we had very low detector efficiency.  Taking into account the low energy of the used  $\gamma$-line, the high Z and the thickness of the used target, correction was made for  $\gamma$-ray self-absorption in the target (around 23\%).
For assessment of the contaminating long-lived $^{160}$Tb many independent  $\gamma$-lines are available (see Table 1).
The uncertainties on the cross-section values are based on the standard approach described in the ISO guide \cite{14} and are obtained as the square route of the quadratic summation of the independently contributing parameters. Uncertainty on the energy scale includes cumulative effects of initial beam energy, stopping estimation influenced by target thickness and beam straggling.

\begin{table}
\tiny
\caption{\textbf{Decay characteristic of the investigated reaction products}}
\label{tab:1}       
\begin{tabular}{|p{0.5in}|p{0.3in}|p{0.5in}|p{0.35in}|p{0.45in}|p{0.35in}|} \hline 
Nuclide & Half-life & E${}_{\gamma}$(keV) & I${}_{g}$${}_{\gamma}$(\%) & Contributing reaction & Q-value\newline (keV) \\ \hline 
\textbf{${}^{1}$${}^{61}$Tb\newline }~$\beta $${}^{-}$: 100 \%\textbf{} & 6.89 d & 25.65135\newline 48.91533\newline 57.1917\newline 74.56669 & ~23.2\newline ~17.0\newline ~1.79\newline 10.2 & ${}^{1}$${}^{60}$Gd(d,n) & 4583.96 \\ \hline 
\textbf{${}^{1}$${}^{61}$Gd\newline }$\beta $${}^{-}$: 100 \%~\textbf{} & 3.66 m & ~102.315\newline ~314.92\newline ~360.94 & 13.9\newline ~22.7\newline 60.1 & ${}^{1}$${}^{60}$Gd(d,p) & 3410.83 \\ \hline 
\textbf{${}^{1}$${}^{60}$Tb\newline }$\beta $${}^{-}$: 100 \%\textbf{} & 72.3 d & 86.7877\newline 197.0341\newline 215.6452\newline 298.5783\newline 879.378\newline 962.311\newline 966.166\newline 1177.954 & 13.2\newline 5.18\newline 4.02\newline 26.1\newline 30.1\newline 9.81\newline 25.1\newline 14.9 & ${}^{1}$${}^{60}$Gd(d,2n) & -3112.6 \\ \hline 
\end{tabular}

\end{table}

\section{Theoretical calculations}
\label{sec:3}

The cross-sections of the investigated reactions were calculated using the pre-compound model codes ALICE-IPPE \cite{15} and EMPIRE-II \cite{16}. To improve the description of the experimental data of deuteron induced reactions at lower energies modified model codes were used.
In the ALICE IPPE-D and EMPIRE-D code versions the direct (d,p) channel is increased strongly, which reflects in the results for all other reactions in both codes \cite{17,18}. A comparison with the data retrieved from the online TENDL 2011 and TENDL 2012 libraries, based on the latest version of the TALYS code system \cite{19} is also made. We presented both TENDL 2011 and 2012 versions in the figures to see the improvement of the description.

\section{Results}
\label{sec:4}
When supposing that highly enriched $^{160}$Gd targets are used, the measured production cross-sections on $^{nat}$Gd can be transformed into reaction cross-sections on pure $^{160}$Gd. From the point of view of radionuclidic purity only the long-lived $^{160}$Tb (T$_{1/2}$=72.3 d) should be considered as a possible and relevant contaminant up to the threshold of the $^{160}$Gd(d,6n)$^{156}$Tb reaction ( Q = -33560.68 keV). Other Tb radioisotopes produced at lower energy are either stable ($^{159}$Tb) or have very long half-life ($^{158}$Tb (150 a) and $^{157}$Tb (150 a)).
The cumulative excitation  function for  the $^{160}$Gd(d,x)$^{161}$Tb reaction is the sum of $^{160}$Gd(d,n)$^{161}$Tb and $^{160}$Gd(d,p)$^{161}$Gd$\rightarrow$ $^{161}$Tb and the excitation function of the $^{160}$Gd(d,2n)$^{160}$Tb reaction, compared to the results of three theoretical codes are presented in Fig. 1 and Fig. 3. The numerical data are presented in Table 2.

\begin{table}
\tiny
\caption{\textbf{Experimental cross-sections for $^{160}$Gd(d,x)$^{161}$Tb(cum) and  $^{160}$Gd(d,x)$^{160}$Tb reactions}}
\label{tab:2}       
\begin{tabular}{|p{0.4in}|p{0.3in}|p{0.4in}|p{0.3in}|p{0.4in}|p{0.3in}|} \hline 
\multicolumn{2}{|c|}{E $\pm\Delta$E\newline (MeV)} & \multicolumn{4}{|c|}{Cross-section($\sigma$)$\pm\Delta\sigma$\newline (mb)} \\ \hline 
 &  & \multicolumn{2}{|p{0.8in}|}{${}^{161}$Tb(cum)} & \multicolumn{2}{|p{0.8in}|}{${}^{160}$Tb} \\ \hline 
\multicolumn{6}{|c|}{\textbf{Series 1}} \\ \hline 
49.2 & 0.3 &  &  & 49.6 & 8.4 \\ \hline 
47.8 & 0.3 &  &  & 52.9 & 13.0 \\ \hline 
46.4 & 0.3 & 76.1 & 15.9 & 52.2 & 6.3 \\ \hline 
45.0 & 0.4 & 66.1 & 36.8 & 57.5 & 14.8 \\ \hline 
43.6 & 0.4 &  &  & 52.2 & 10.1 \\ \hline 
42.1 & 0.4 &  &  & 45.4 & 11.5 \\ \hline 
40.5 & 0.4 & 77.0 & 8.3 & 48.7 & 12.4 \\ \hline 
39.0 & 0.5 &  &  & 65.8 & 14.2 \\ \hline 
37.4 & 0.5 & 72.2 & 29.0 & 72.4 & 13.6 \\ \hline 
35.7 & 0.5 & 68.4 & 7.4 & 74.2 & 19.1 \\ \hline 
34.0 & 0.5 &  &  & 100.0 & 17.8 \\ \hline 
32.2 & 0.6 &  &  & 79.6 & 15.0 \\ \hline 
30.4 & 0.6 & 105.2 & 11.3 & 109.4 & 17.3 \\ \hline 
28.4 & 0.6 & 104.2 & 11.2 & 107.8 & 16.8 \\ \hline 
26.4 & 0.7 & 112.5 & 39.5 & 105.5 & 16.2 \\ \hline 
24.2 & 0.7 & 147.0 & 15.9 & 132.4 & 18.4 \\ \hline 
21.9 & 0.7 & 159.9 & 17.3 & 147.1 & 19.8 \\ \hline 
19.4 & 0.8 & 192.5 & 20.9 & 197.2 & 23.3 \\ \hline 
16.6 & 0.8 & 203.7 & 22.1 & 329.4 & 37.4 \\ \hline 
13.5 & 0.9 & 223.1 & 24.2 & 553.1 & 60.8 \\ \hline 
\multicolumn{6}{|c|}{\textbf{Series 2}} \\ \hline 
19.8 & 0.3 & 144.8 & 26.0 & 180.3 & 20.7 \\ \hline 
18.3 & 0.3 & 165.6 & 22.3 & 239.0 & 26.9 \\ \hline 
16.6 & 0.4 & 176.4 & 22.9 & 334.6 & 37.2 \\ \hline 
14.9 & 0.4 & 180.3 & 21.0 & 432.1 & 47.1 \\ \hline 
13.0 & 0.4 & 249.7 & 28.8 & 663.7 & 72.3 \\ \hline 
10.9 & 0.5 & 292.9 & 33.4 & 544.0 & 59.9 \\ \hline 
8.4 & 0.5 & 174.0 & 19.6 & 378.9 & 41.6 \\ \hline 
5.4 & 0.6 & 5.1 & 0.9 & 14.1 & 1.8 \\ \hline 
\end{tabular}
\end{table}

The uncertainties on $^{161}$Tb cross-sections are large (and the data are scattered), due to the low statistics of the 57.2 keV  $\gamma$-line and to the uncertainty of the peak fitting in a complex spectrum. The difference between the particular uncertainties can be explained by the fact that different spectra were measured after different cooling times and for different measuring times, so some of them have worse statistics than others.

\begin{figure}
  \includegraphics[width=0.5\textwidth]{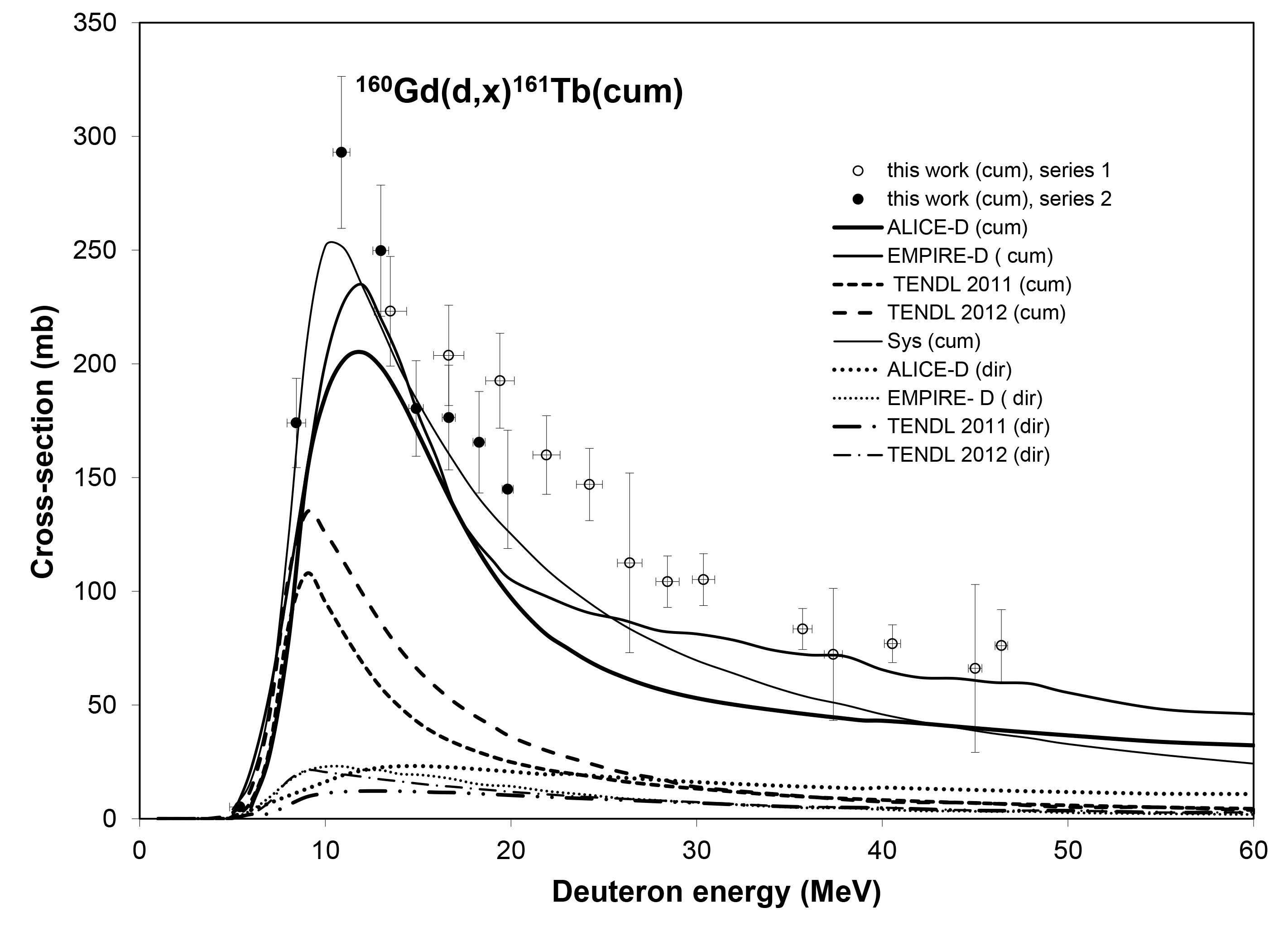}
\caption{Excitation function for the reactions $^{160}$Gd(d,x)$^{161}$Tb(cum) and comparison with the theoretical calculation }
\label{fig:1}       
\end{figure}

In Fig. 1 we have included the results for cumulative production of $^{161}$Tb  using the phenomenological systematics of (d,p)  reaction taken from contribution of A.V. Ignatyuk to FENDL-3 Activation Library \cite{20} (Fig. 1, sys)
The most striking information that can be derived from the Fig. 1 is:
\begin{itemize}
\item	The direct contribution to $^{161}$Tb production is very low.  It is based on the well-known systematics of the (d,n) reactions in this mass region and from our theoretical calculations.
\item	The $^{161}$Tb is produced mostly indirectly.
\item	The agreement with the results of the modified ALICE and EMPIRE codes is acceptable good.
\item	Both TENDL 2011 and TENDL 2012 underestimate significantly the cumulative cross-section (and consequently the (d,p) by a factor of 2). The agreement with the TENDL 2012 results is slightly better. 
\end{itemize}

\begin{figure}
\includegraphics[width=0.5\textwidth]{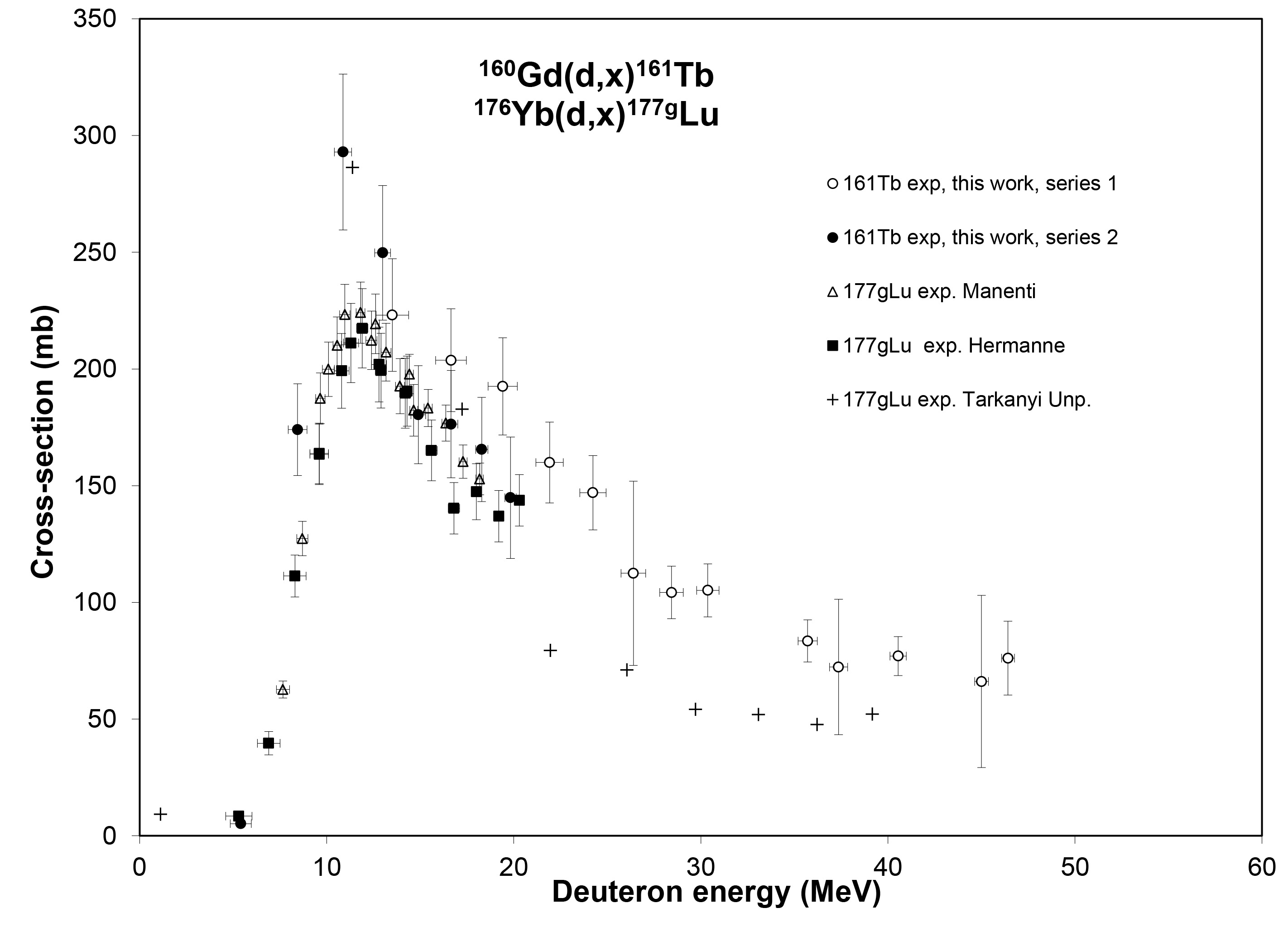}

\caption{Comparison of the excitation functions for the reactions $^{160}$Gd(d,x)$^{161}$Tb(cum) and and $^{176}$Yb(d,x)$^{177g}$Lu(cum) }
\label{fig:2}       
\end{figure}

The cumulative excitation functions of the $^{160}$Gd(d,x)\-$^{161}$Tb and the $^{176}$Yb(d,x)$^{177g}$Lu \cite{2,3,21} are similar both in shape and in magnitude as it is illustrated in Fig. 2.  (The direct and indirect production of the $^{177m}$Lu is negligible due to the high spin value J$^\pi$ = 23/2$^-$). In Fig. 2 our new data (preliminary, unpublished) measured for $^{176}$Yb(d,x)$^{177g}$Lu  reaction up to 40 MeV are also shown for comparison at the high energy range.
The cumulative excitation function for the $^{160}$Gd(d,2n)\-$^{160}$Tb reaction is shown in Fig. 3  in comparison with the results of the model codes and the following conclusions can be drawn:
\begin{itemize}
\item	The contaminating $^{160}$Tb is produced directly via $^{160}$Gd(d,2n) reaction.
\item	The agreement with the results of the modified ALICE-D and TENDL 2011 and TENDL 2012 is acceptable.
\item	The results of EMPIRE-D overestimate the experimental values. 
\item	 At high energies above 20 MeV the results of all model codes are systematically too low.
\item	The maximum of the (d,2n) cross-section in this mass region is high, in agreement with the systematics.
\end{itemize}

\begin{figure}
\includegraphics[width=0.5\textwidth]{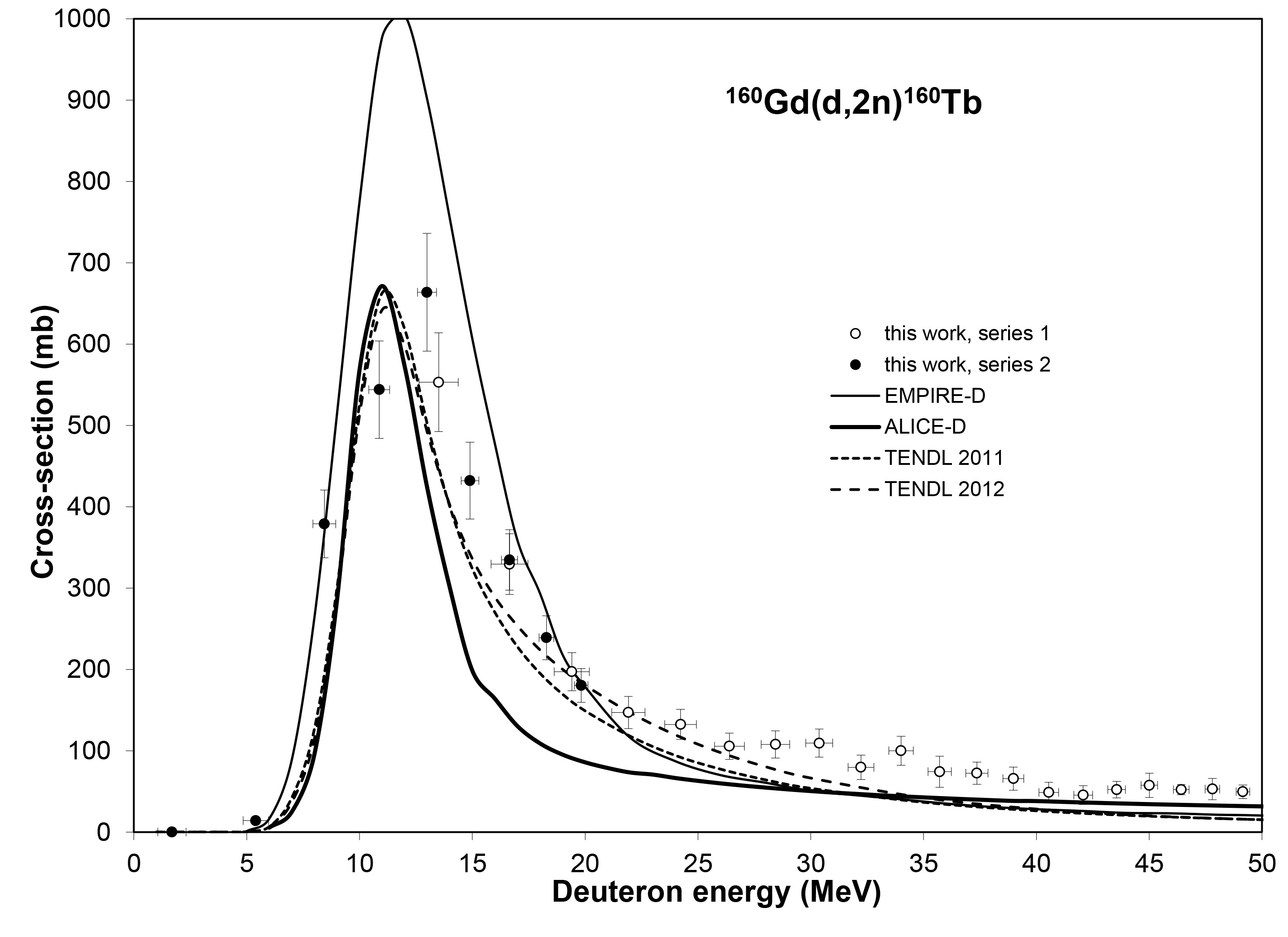}
\caption{Excitation function for the reactions $^{160}$Gd(d,2n)$^{160}$Tb and comparison with the theoretical results}
\label{fig:3}       
\end{figure}
 
\section{Thick target yields}
\label{sec:5}
From fitted curves to our experimental cross-sections integral thick target yields were calculated for production of $^{161}$Tb(cum) and $^{160}$Tb. The results for physical yields \cite{22} are presented in Fig. 4.  No earlier experimental thick target yield data were found in the literature for comparison.

\begin{figure}
\includegraphics[width=0.5\textwidth]{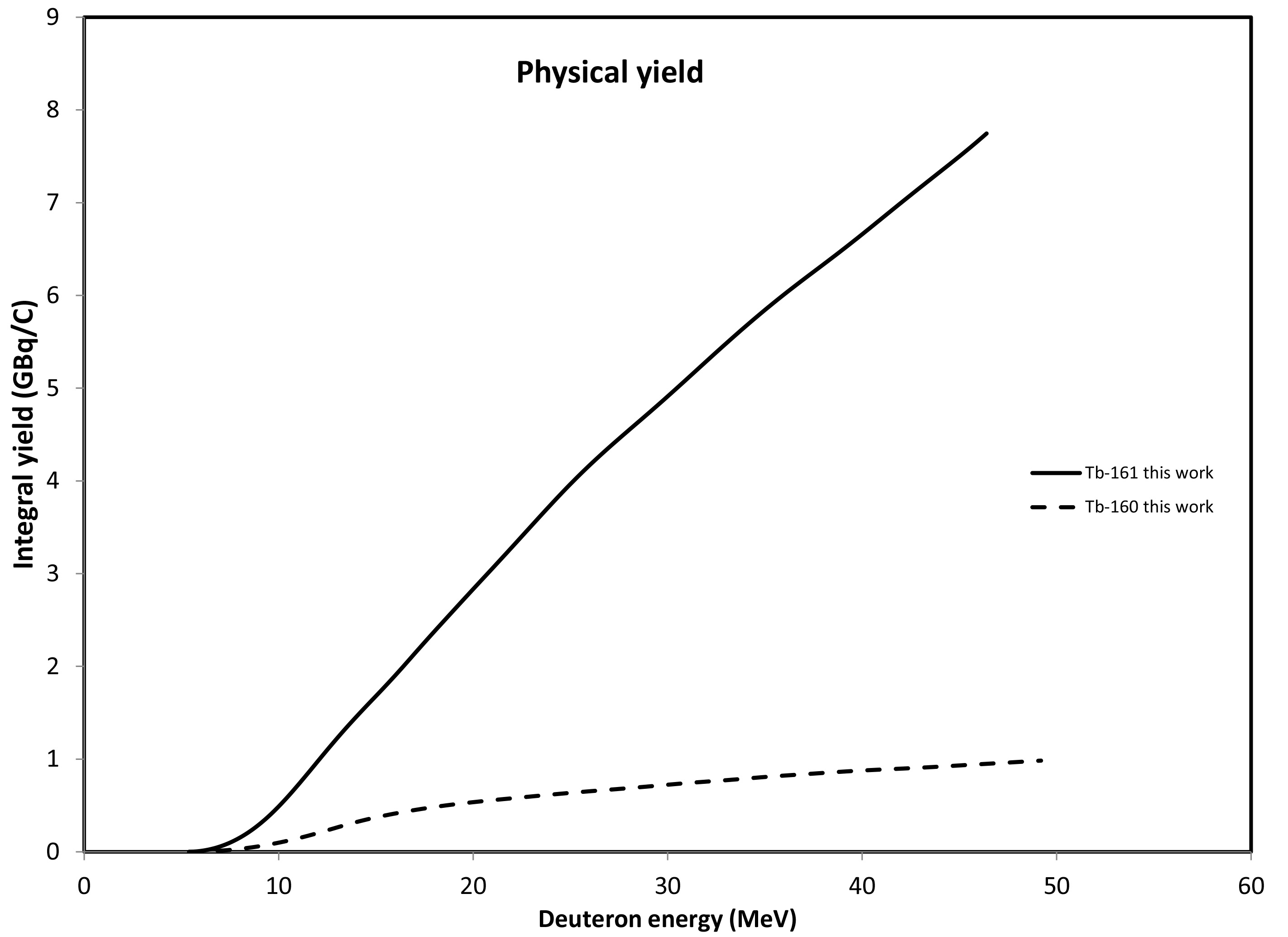}
\caption{Integral yields for $^{160}$Gd(d,x)$^{161}$Tb and $^{160}$Gd(d,x)$^{160}$Tb  reactions}
\label{fig:4}       
\end{figure}

\section{Comparison of the accelerator and reactor production routes}
\label{sec:6}
The presently used route for production of carrier free $^{161}$Tb is the indirect $^{160}$Gd(n,?)$^{161}$Gd$\rightarrow$ $^{161}$Tb  process at high flux nuclear research reactors or at spallation neutron sources \cite{1,23} followed by a chemical separation of radionuclide of interest from the target material. According to the present study the $^{161}$Tb can be produced also at accelerators. The main parameters for comparison of the reactor and accelerator routes are collected in Table 3.

\begin{table}
\tiny
\caption{\textbf{Comparison of production routes of ${}^{161}$Tb}}
\label{tab:3}       
\begin{tabular}{|p{0.8in}|p{0.5in}|p{0.4in}|p{0.55in}|p{0.55in}|} \hline 
 & \multicolumn{2}{|c|}{Reactor} & \multicolumn{2}{|c|}{Accelerator} \\ \hline 
Reaction & \multicolumn{2}{|c|}{${}^{160}$Gd(n,$\gamma $)${}^{161}$Gd$\rightarrow$${}^{161}$Tb  } & \multicolumn{2}{|c|}{${}^{160}$Gd(d,x)${}^{161}$Tb} \\ \hline 
Flux (part/sec cm${}^{2}$) & 1.5*10${}^{15}$ & 3*10${}^{13}$ & {1.53*10${}^{15 }$p/sec\newline (250 mA)\newline } & 6.1*10${}^{15}$p/sec\newline (1000 mA)\newline  \\ \hline 
Energy range  & thermal & thermal & {30-5 MeV} & 30-5 MeV \\ \hline 
Target mass (mg) & 40 & 40 & {3} & 3 \\ \hline 
Average cross-section (mb) & 1500 & 1500 & {125} & 125 \\ \hline 
Yield ratio to high flux reactor & 1 & 1/50 & {1/266} & 1/66.5 \\ \hline 
Irradiation & \multicolumn{2}{|p{0.9in}|}{multichannel} & \multicolumn{2}{|p{1.1in}|}{single channel} \\ \hline 
Chemical separation & \multicolumn{2}{|p{0.9in}|}{yes} & \multicolumn{2}{|p{1.1in}|}{yes} \\ \hline 
Enrichment  & \multicolumn{2}{|p{0.9in}|}{highly enriched} & \multicolumn{2}{|p{1.1in}|}{highly enriched} \\ \hline 
Specific activity & \multicolumn{2}{|p{1.in}|}{higher} &  & {} \\ \hline 
Disturbing reaction  &  &  & \multicolumn{2}{|p{0.9in}|}{${}^{160}$Gd(d,2n)${}^{160}$Tb\newline } \\ \hline 
\end{tabular}
\end{table}

\subsection{Production yield}
\label{6.1}
The Table 3 shows that from the point of view of production yield of $^{161}$Tb the recent accelerators are not competitive with a high flux reactor. The productivity of a high intensity accelerator is close to the productivity of standard reactors.

\subsection{Targetry}
\label{6.2} 
Use of charged particle beam requires special, high power targetry, which is not a simple question. The thermal conductivity of Tb is 40 times lower than that of Cu (Tb-10.6 W·m$^1$·K$^1$   and   Cu-400 W·m$^1$·K$^1$). 
Using highly enriched target material, different target preparation methods can be considered. Electrodeposited in metal form can be used taking into account that the Gd metal is relatively stable in dry air \cite{24}. Different compounds (oxide, chloride) can be used, prepared by pressing or in molten form. 

\subsection{Radionuclidic purity, specific activity}
\label{6.3}
In case of neutrons, in principle there is no disturbing impurity by using highly enriched targets followed by an effective chemical separation. 
By using deuterons up to 40 MeV of commercial cyclotrons on $^{nat}$Gd targets a considerable amount of $^{153,154,155,156,157,158,160}$Tb will be produced. Therefore the radionuclidic impurity will be high.
By using highly enriched $^{160}$Gd (100\%) up to 20 MeV bombarding energy, only $^{160}$Tb  and $^{161}$Tb radioisotopes will be produced. The energy domain of the excitation functions of the $^{161}$Tb and $^{160}$Tb is nearly similar but the cross-section is 2-3 times higher for the production of $^{160}$Tb. Therefore it is difficult to reduce the radionuclidic impurities and to increase the specific activity by using an adapted energy domain (see Fig. 5). Roughly two times more $^{160}$Tb than $^{161}$Tb nuclei will be present in the irradiated sample, even using short irradiations (see Fig. 5). The ratio will even be worse for longer irradiations (more than 100 h), because it will increase with decay after EOB approximately linearly. Because the allowed radionuclidic impurity may not exceed 1-2\% in the case of human application, this way of production is completely unsuitable for medical purposes. 

\begin{figure}
\includegraphics[width=0.5\textwidth]{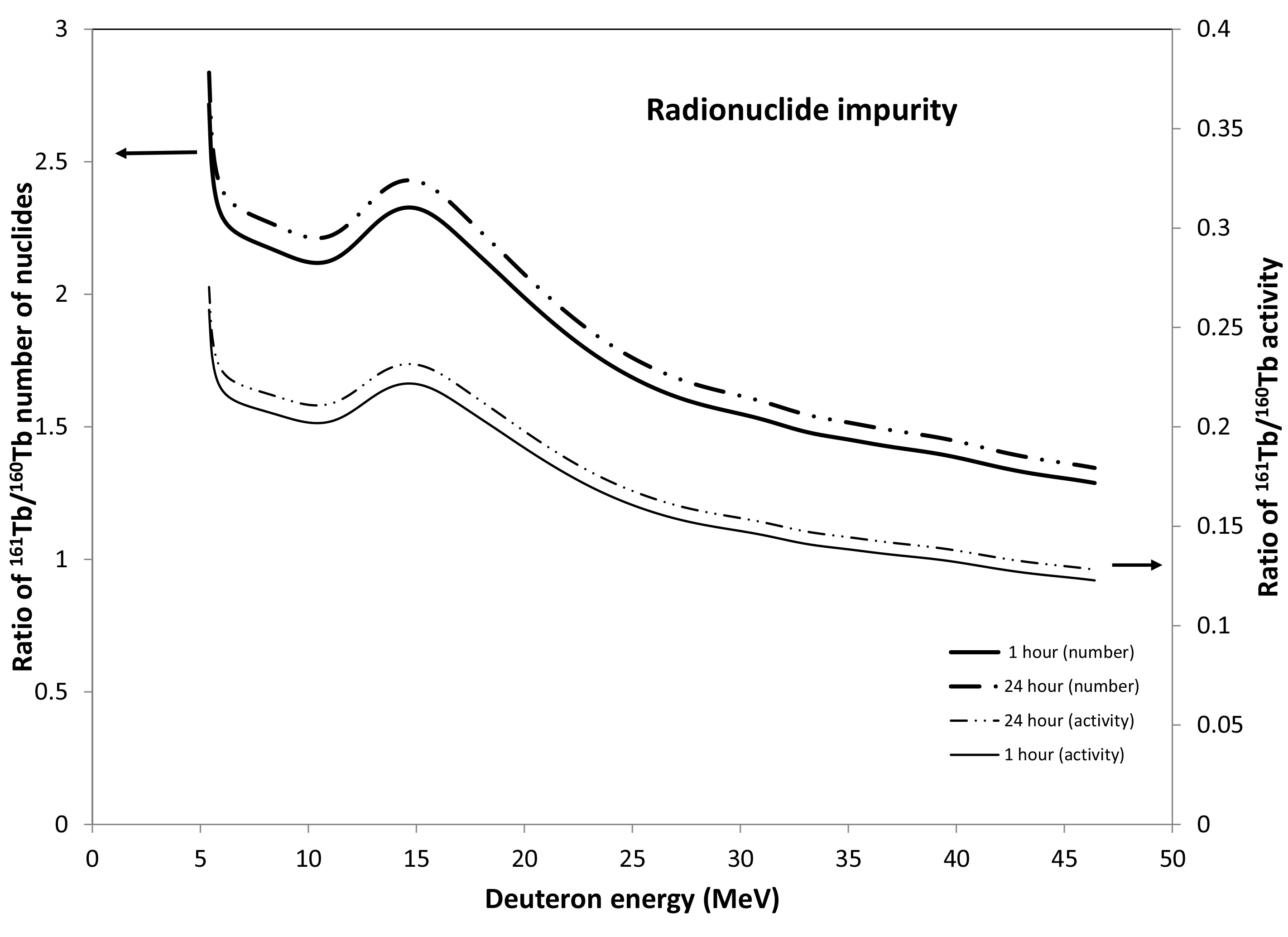}
\caption{5.	Radionuclidic impurity of the irradiated highly enriched $^{160}$Gd after different irradiation times. Both the ratio of the number of radionuclides (left) and the ratio of activities (right) are presented.}
\label{fig:5}       
\end{figure}

The conclusion based on the short discussion above is that accelerator production of $^{161}$Tb is not competitive with the reactor production both in the aspect of commercially attractiveness (investment in irradiation facility not considered) and product quality.  The only advantage of ion-accelerators production could be the easier availability in small batches for research purposes.

\section{Comparison of the experimental and theoretical cross-section data}
\label{7}
To model a reliable and accurate theoretical description of deuteron induced activation mechanisms knowledge of experimental data is very important. The interaction of the deuteron with the target nuclei is very complex, as it includes the deuteron break up process and the related stripping and pick-up mechanisms, statistical emission, etc. Possibilities to take these different processes into account in model codes have been proposed in different-simple and complex ways \cite{18,25,26}.
The comparison with the present experimental data also shows that the theoretical description still needs significant improvement in the TALYS code, especially for (d,pxn) processes. In case of the modified ALICE and EMPIRE codes the agreement is better, but it is a result of a correction based on a phenomenological simulation of direct (d,p) and (d,t) transitions,  including a "reduction  factor" of the compound nucleus cross-section due to direct processes based on  systematic behavior of the experimental data.
Our systematical study of the deuteron induced reactions \cite{27,28} can contribute effectively to further improvement of the theoretical treatment.

\begin{acknowledgements}
This study was performed in the frame of the MTA-JSPS and MTA-FWO (Vlaanderen) collaboration programs. The authors thank the different research projects and their respective institutions for the practical help and providing the use of the facilities for this study.
\end{acknowledgements}

\bibliographystyle{spphys}       
\bibliography{Gdd}   

%
%

\end{document}